# Evaluation of interval extension of the power function by graph decomposition

Evgueni Petrov, pes@iis.nsk.su

A.P. Ershov Institute of Informatics Systems
prospekt akademika Lavrentjeva, dom 6, Novosibirsk 630090, RUSSIA

**Abstract**

The subject of our talk is the correct evaluation of interval extension of the function specified by the expression $x^y$.

Despite the fact that the algorithms which calculate interval extensions of the basic mathematical functions are well known (the algorithms which do the arithmetic operations on intervals are even patented in the USA), at present, there is no systematic approach to evaluation of interval extension of the power function.

The existing packages of mathematical software for interval calculations allow one to calculate interval extension of the power function but constrain the values of the variables $x$ и $y$. The most common constraints are $x \in R$, $y \in Z$ and $x \in R_+ = [0, \infty)$, $y \in R$. The presence of such constraints complicates the use of interval software for reliable solution of mathematical modeling problems. First, these artificial constraints complicate solution of problems with unknown power exponent. Second, these constraints complicate the use of interval software in cooperation with symbolic preprocessors of mathematical models. For example, replacing $x^3$=const with $x$=const$^{1/3}$ is not a valid symbolic transformation for the power function defined only for positive bases.

Most likely, such constraints on evaluation of the power function emerge from a cautious attitude of the developers of interval software to discontinuous functions: the power function, as a function of two variables, is discontinuous (or undefined) at every point $(x, y)$ where the base $x$ is negative. In fact, the constraints artificially set for the values of $x$ and $y$, specify the domains where the power function is continuous.

The goal for the research which is the subject of our talk was development of a systematic approach to the correct evaluation of interval extension of the power function $x^y$ without any constraints on the values of $x$ and $y$. The approach which we have developed is used for implementation of the power function in the UniCalc system [1].

The core of our approach is a decomposition of the graph of the function $x^y$ into a small number of parts which can be transformed, by reflections with respect to coordinate planes, into subsets of the graph of $x^y$ for non-negative bases $x$. By construction, the closure of each such subset is the entire graph of $x^y$ (in the topology of Euclidean space $R^3$). Because of this fact, evaluation of interval extension of $x^y$ without any constraints on the values of $x$ and $y$, is not much harder than evaluation of interval extension of $x^y$ for non-negative bases $x$.

We consider $x^y$ as a restriction of the power function of complex variables calculated for the principal branch of the argument $\arg(x) \in [0, 2\cdot\pi)$: $x^y = |x|^y \cdot (\cos(y\cdot\arg(x)) + i\cdot\sin(y\cdot\arg(x)))$ for $x \neq 0$, $0^y = 0$ for positive $y$, $0^0 = 1$. It is easy to see that the function specified this way takes real values (i.e., $\sin(y\cdot\arg(x))=0$) only in the following cases:
- $y$ is an irreducible fraction with odd denominator, $x \in R$;
- $y$ is an irreducible fraction with even denominator (and odd numerator), $x \in R_+$;
- $y$ is an irrational number, $x \in R_+$.

We decompose the graph of $x^y$ into the following parts:
- $y$ is an irreducible fraction with odd denominator and even numerator;
- $y$ is an irreducible fraction with odd denominator and numerator;
- $y$ is an irreducible fraction with even denominator, $x \in R_+$;
- $y$ is an irrational number, $x \in R_+$.

Denote by $X^Y$ the image of the Cartesian product of the sets (not necessarily intervals) $X \cap R_+$ and $Y$ under the power function. Denote by $Y_{eo}$, $Y_{oo}$, $Y_{oe}$, $Y_{ir}$ the subsets of $Y$ which consist, respectively, of the irreducible fractions with odd denominator and even numerator, odd denominator and numerator, even numerator, and irrational numbers.

According to our decomposition of the entire graph of $x^y$, the image of the box $(\mathbf{x}, \mathbf{y})$ under the power function is $\mathbf{x}^{\mathbf{y}_{eo}} \cup (-\mathbf{x})^{\mathbf{y}_{eo}} \cup \mathbf{x}^{\mathbf{y}_{oo}} \cup -(-\mathbf{x})^{\mathbf{y}_{oo}} \cup \mathbf{x}^{\mathbf{y}_{oe}} \cup \mathbf{x}^{\mathbf{y}_{ir}}$. After an obvious reduction, this expression becomes $\mathbf{x}^\mathbf{y} \cup (-\mathbf{x})^{\mathbf{y}_{eo}} \cup -(-\mathbf{x})^{\mathbf{y}_{oo}}$. Note that, if $\mathbf{y}$ is a non-singleton interval, then the closure of the sets $\mathbf{y}_{eo}$ and $\mathbf{y}_{oo}$ is the entire interval $\mathbf{y}$.

Denote by $\mathbf{p_0}(\mathbf{x}, \mathbf{y})$ interval extension of the power function $x^y$ for non-negative base. Taking into account the considerations from the previous paragraph, interval extension $\mathbf{p}(\mathbf{x}, \mathbf{y})$ of the power function, without constraints on the values of $x$ and $y$, is expressed in terms $\mathbf{p_0}$ of as follows:

- $\mathbf{p}(\mathbf{x}, \mathbf{y}) = \mathbf{p_0}(\mathbf{x}, \mathbf{y}) \cup \mathbf{p_0}(-\mathbf{x}, \mathbf{y}) \cup -\mathbf{p_0}(-\mathbf{x}, \mathbf{y})$, if $\mathbf{y}$ is a non-singleton interval ;
- $\mathbf{p}(\mathbf{x}, \mathbf{y}) = \mathbf{p_0}(\mathbf{x}, \mathbf{y}) \cup \mathbf{p_0}(-\mathbf{x}, \mathbf{y})$, if $\mathbf{y}$ is a singleton interval which contains an irreducible fraction of the form "even/odd";
- $\mathbf{p}(\mathbf{x}, \mathbf{y}) = \mathbf{p_0}(\mathbf{x}, \mathbf{y}) \cup -\mathbf{p_0}(-\mathbf{x}, \mathbf{y})$, if $\mathbf{y}$ is a singleton interval which contains an irreducible fraction of the form "odd/odd";
- $\mathbf{p}(\mathbf{x}, \mathbf{y}) = \mathbf{p_0}(\mathbf{x}, \mathbf{y})$, if $\mathbf{y}$ is a singleton interval which contains an irrational number.

Of practical interest is the restriction $\mathbf{p_f}(\mathbf{x}, \mathbf{y})$ of interval extension $\mathbf{p}(\mathbf{x}, \mathbf{y})$ of the power function to the set of intervals having computer representable bounds. Because the set of computer representable real numbers does not contain any irrational numbers or irreducible fractions with an odd denominator different from 1 and –1, the following equations hold:

- $\mathbf{p_f}(\mathbf{x}, \mathbf{y}) = \mathbf{p_0}(\mathbf{x}, \mathbf{y}) \cup \mathbf{p_0}(-\mathbf{x}, \mathbf{y}) \cup -\mathbf{p_0}(-\mathbf{x}, \mathbf{y})$, if $\mathbf{y}$ is a non-singleton interval;
- $\mathbf{p_f}(\mathbf{x}, \mathbf{y}) = \mathbf{p_0}(\mathbf{x}, \mathbf{y}) \cup \mathbf{p_0}(-\mathbf{x}, \mathbf{y})$, if $\mathbf{y}$ is a singleton interval containing an even integer;
- $\mathbf{p_f}(\mathbf{x}, \mathbf{y}) = \mathbf{p_0}(\mathbf{x}, \mathbf{y}) \cup -\mathbf{p_0}(-\mathbf{x}, \mathbf{y})$, if $\mathbf{y}$ is a singleton interval containing an odd integer.